\documentclass{elsart1p}
\usepackage{graphicx}
\usepackage{amssymb}
\begin{document}
\begin{frontmatter}
\date{7 January 2011}

\title{Performance studies of the PHENIX Hadron Blind Detector at RHIC}

\author{P. Garg} and {B. K. Singh}
\\(For PHENIX Collaboration)
\address{Department of Physics, Banaras Hindu University, Varanasi-221005, India}

\begin{abstract}
Electron pairs or di-leptons in general are unique probes to study the
hot and dense matter formed in relativistic heavy ion collisions at RHIC. Particularly, low mass di-leptons are sensitive to chiral symmetry restoration effects and to thermal radiation emitted by the plasma via virtual photons, providing a direct measurement of the quark gluon plasma temperature. But the experimental challenge is the huge combinatorial background created by $e^+e^-$ pairs from copiously produced $\pi^0$ Dalitz decay and $\gamma$ conversions. In order to reduce this background, a Hadron Blind Detector was proposed in PHENIX
for electron identification in high-density hadron environment. In the present paper some of the performance studies of the HBD carried with data from 2009 Run are discussed.
\end{abstract}

\begin{keyword}Quark Gluon Plasma, Low mass Dileptons, Hadron Blind Detector
\end{keyword}
\end{frontmatter}

\section{Introduction}

Hadron Blind Detector (HBD) was developed and installed as an upgrade of the PHENIX experiment at the Relativistic Heavy Ion Collider (RHIC) experiment for the measurement of electron pairs, particularly in the low mass region ($m\leq$ 1 GeV/$c^2$ including the light vector mesons $\rho$, $\omega$ and $\phi$)[1]. Dileptons are valuable probes for the hot and dense matter formed in ultra-relativistic heavy-ion collisions. They play a crucial role in the quest for the QCD phase transition from hadron gas (HG) to the quark gluon plasma (QGP) expected to be formed in these collisions. They can provide evidence of chiral symmetry restoration and deconfinement phase transition[2].
\\
Dileptons offer also the possibility to identify the thermal radiation emitted from the QGP via $q \overline{q}$ annihilation. Such a radiation is regarded as a very strong signal of deconfinement. There is no convincing evidence for thermal radiation
from the QGP at lower energies, either in the dilepton or in the real
photon channels. Theoretical calculations have singled out the
dilepton intermediate mass range $(m = 1-3 GeV/c^2)$ as the most
appropriate window for the observation of QGP thermal radiation[4].
\\
The measurement of dileptons in heavy ion collisions is a  challenging task because of the huge combinatorial background in the low mass region. Electron pairs measured in PHENIX from the 2004 Run showed very low signal-to-background ratio S/B  $\sim$ 1/200 in the low mass region with large statistical and systematic uncertainties (Fig.1).
\begin{figure}
\begin{center}
\includegraphics[scale=0.35]{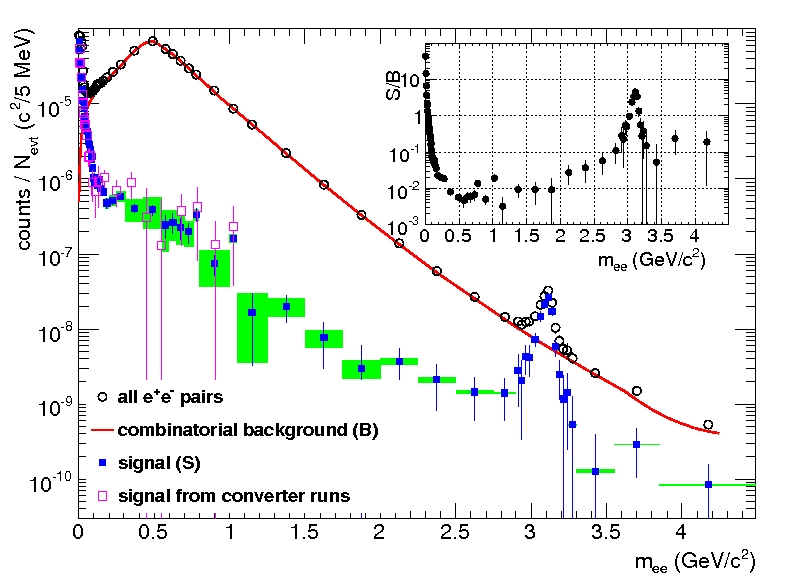}
\caption{Background(B) and Signal(S) with statistical(bar) and systematic(boxes) uncertainties. The inset shows S/B ratio[3].}
\end{center}
\end{figure}
Moreover, the limited azimuthal angular acceptance in the central arms and the strong magnetic field beginning radially at R=0, makes the identification and rejection of electron-positron pairs from Dalitz decays and photon conversions very difficult. Therefore, the improved S/B is needed for further studies of low mass dilepton spectra. The main sources of combinatorial background are coming from $\pi^0\rightarrow\gamma\; e^{+}e^{-}$ and  $\pi^{0}\rightarrow\gamma\gamma\rightarrow\gamma\; e^{+} e^{-}$, and we exploited the fact that these channels have small opening angle so by using the opening angle we must be able to distinguish single hits from double hits in HBD. 

\label{}

\section{Hadron Blind Detector and Performance Studies}
\subsection{Detector concept}
The main task of HBD was to recognize and reject $\gamma$-conversions and $\pi^0$
Dalitz decay pairs using the fact of their small opening angle. In
order to conserve the opening angle of the decay pairs, HBD was placed in the magnetic
field free region as is shown in figure 2(b). 
HBD consists of a Cherenkov radiator operated with pure CF$_4$ in a proximity focus configuration directly coupled  to a triple-GEM detector element with a CsI photocathode on the top GEM[5] and a pad readout.\\
The choice of CF$_4$ both as radiator and detector gas in a windowless geometry results in a very large bandwidth (from $\sim$6 eV given by the threshold of
the CsI to $\sim$11.5 eV given by the CF$_4$ cut-off) and
consequently in a large figure of merit N$_0$ and a large number of photo electrons N$_{pe}$.
One of the important advantage of using GEMs is that it allows the use of a
reflective photocathode which is totally screened from photons produced in the avalanche process.
HBD contains two arms in east and west side of PHENIX co-ordinate system and its acceptance in pseudorapidity is $|\eta|\leq$ 0.45 and in azimuthal angle ($\Delta\Phi$)  is 135$^0$.
\begin{figure}
\begin{center}
\includegraphics[scale=0.5]{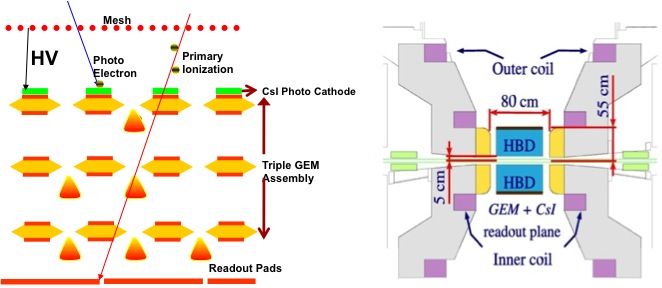}
\caption{Schematic of (a) HBD working principle and (b)  position of HBD in PHENIX.}
\end{center}
\end{figure}

\subsection{Performance}

HBD position resolution is estimated by the matching of the Central Arm tracks to the HBD. The matching distributions (difference between projection point and the closest cluster in HBD) of electron tracks in $\Phi$ and Z were determined for different track momentum bins. In Fig. 3, the momentum dependence of the sigma of $\Delta$Z and $\Delta\Phi$ is shown and the asymptotic resolution in Z and $\Phi$ are obtained $\sim1.05$ cm and 8 mrad respectively.
\begin{figure}
\begin{center}
\vspace{0.6in}
\includegraphics[scale=0.45]{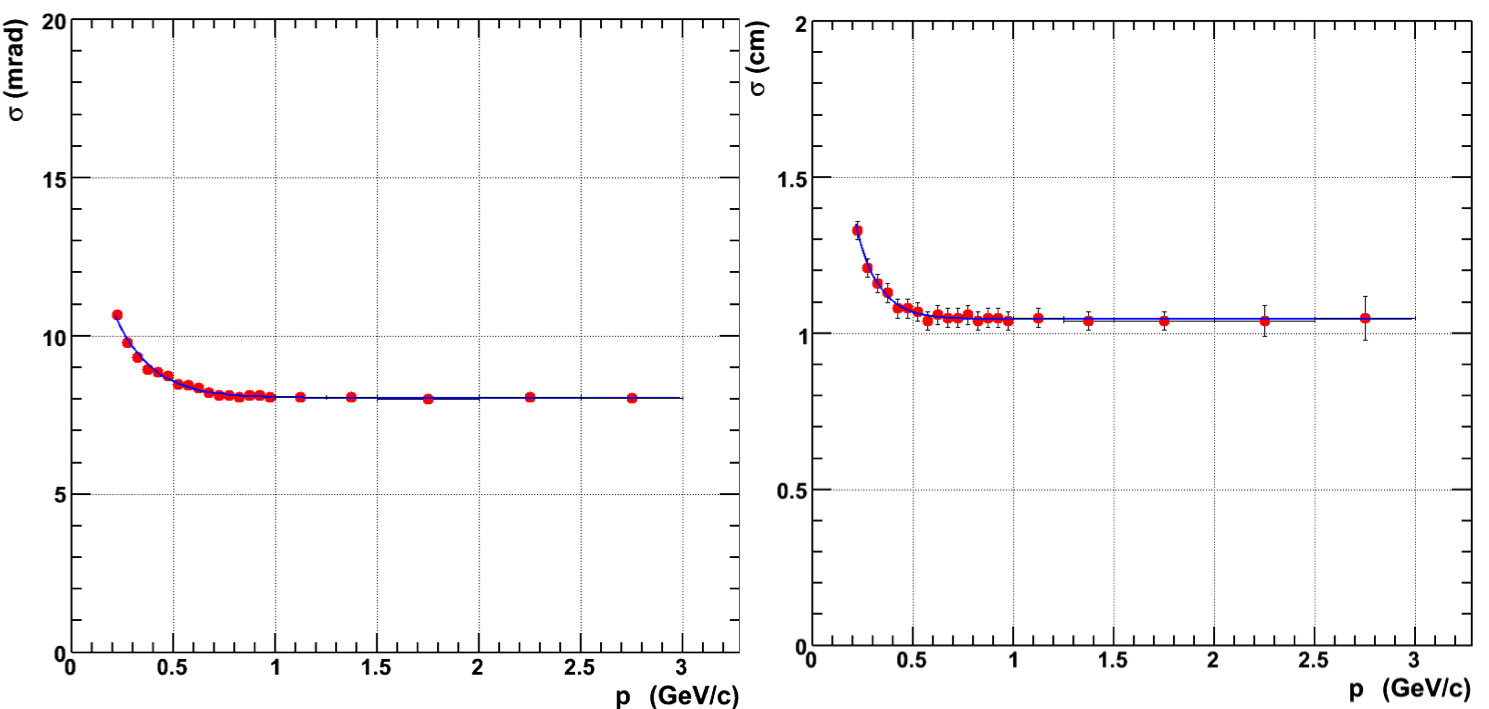}
\caption{The variation of matching resolution of the electron tracks in Z and $\Phi$  with respect to momentum.}
\end{center}
\end{figure}
In Fig.4(a), a drop in pulse height of hadrons in RB mode of HBD is demonstrated. Hadron rejection factor (ratio of the number of hadron tracks identified in central arm detectors to the number of corresponding matched hits in the HBD) as a function of threshold charge in reverse bias mode of HBD for a single module is shown in Fig. 4(b).
Also a very good separation between single electrons and hadrons in RB
has been observed as single electron response peaked at $\sim$ 20
photoelectrons whereas electrons from hadrons are peaked at $\sim$40 due to single
and double hits in HBD respectively.
 Single electron efficiency for the full HBD extracted from di-electrons in the  J/$\Psi$
 mass region is obtained approximately as $90\%$.

\section{Conclusion}
Hadron Blind Detector performed well in PHENIX during RHIC Runs in 2009 ($p$+$p$) and
2010 (Au+Au). The performance studies carried with $p$+$p$ data, the hadron rejection power achieved, good separation between electrons coming from resonances or heavy quarks and neutral hadrons are consequent improvement in the di-electron S/B. The studies of Au+Au data set are in progress.  

\section{Acknowledgements}
\begin{figure}
\begin{center}
\includegraphics[scale=0.3]{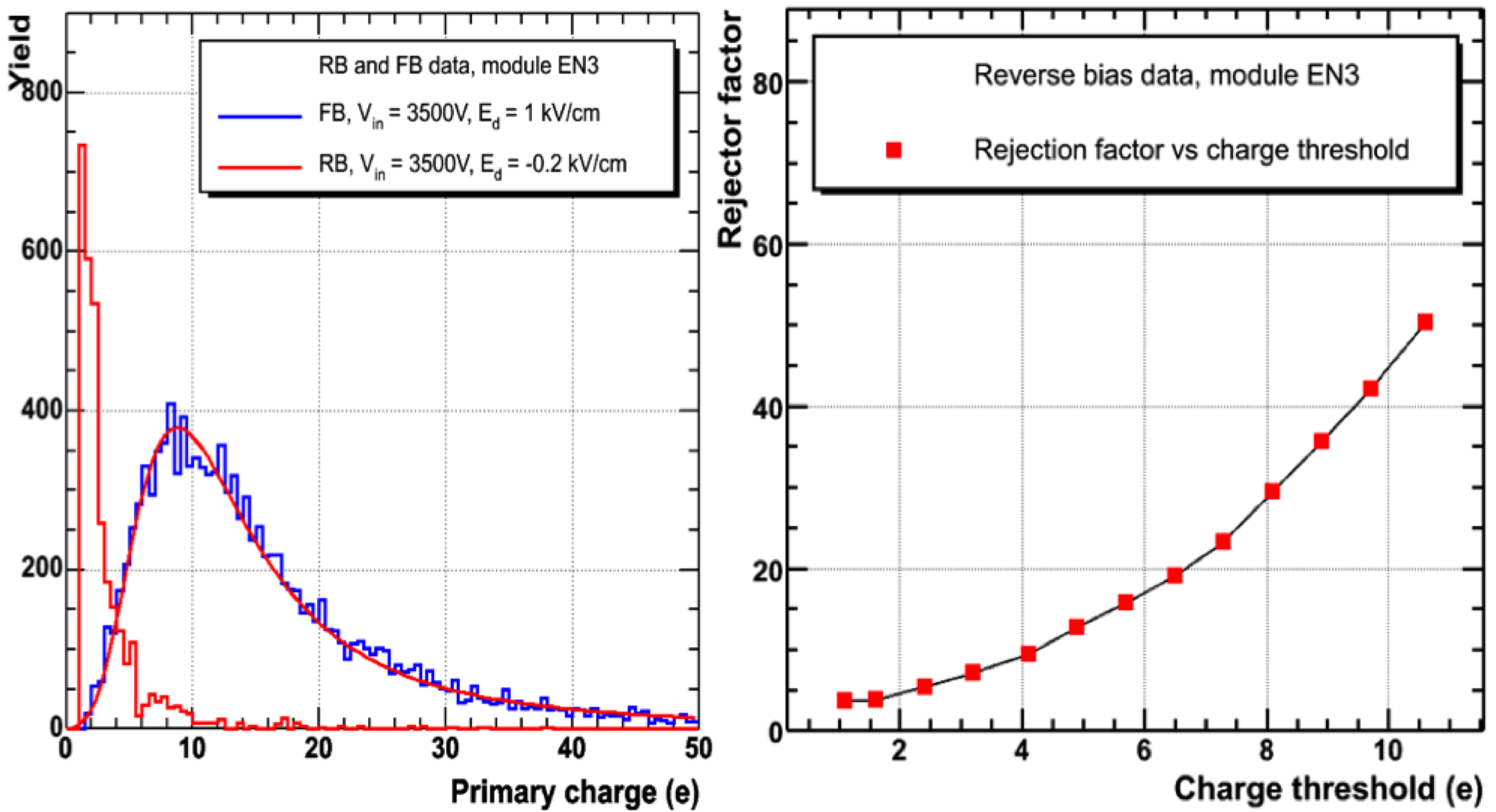}
\caption{(a) Hadron charge spectra in reverse(red) and forward bias(blue) (b) Hadron Rejection Factor as a function of charge threshold.}
\end{center}
\end{figure}
\vskip0.1cm  Authors acknowledge the financial support from Department of Science and Technology (DST), Government of India, New Delhi.

\end{document}